\documentclass[amsmath,amssymb,aps,prl,twocolumn,showpacs,superscriptaddress]{revtex4-1}
\usepackage{graphicx}
\usepackage{dcolumn}
\usepackage{bm}
\usepackage{xr}
\makeatletter
\usepackage{hyperref}
\usepackage{xcolor}
\hypersetup{
  colorlinks   = true,
  urlcolor     = black,
  linkcolor    = blue,
  citecolor   = blue
}

\begin{document}
\title{A bending fluctuation-based mechanism for particle detection by ciliated structures}

\author{Jean-Baptiste Thomazo}
\affiliation{Sorbonne Universit\'e, CNRS, Institut de Biologie Paris-Seine (IBPS), Laboratoire Jean Perrin (LJP), 4 place Jussieu, F-75005 Paris, France.}
\affiliation{Nestl\'e Development Centre Lisieux, rue d'Orival, F-14100 Lisieux, France.}
\author{Benjamin Le R\'ev\'erend} 
\affiliation{Nestl\'e Research, Route du Jorat, CH-1000 Lausanne, Switzerland.}

\author{L\'ea-Laetitia Pontani}
\affiliation{Sorbonne Universit\'e, CNRS, Institut de Biologie Paris-Seine (IBPS), Laboratoire Jean Perrin (LJP), 4 place Jussieu, F-75005 Paris, France.}
\author{Alexis M. Prevost}
\email[]{alexis.prevost@sorbonne-universite.fr}
\affiliation{Sorbonne Universit\'e, CNRS, Institut de Biologie Paris-Seine (IBPS), Laboratoire Jean Perrin (LJP), 4 place Jussieu, F-75005 Paris, France.}

\author{Elie Wandersman}
\email[]{ elie.wandersman@sorbonne-universite.fr}
\affiliation{Sorbonne Universit\'e, CNRS, Institut de Biologie Paris-Seine (IBPS), Laboratoire Jean Perrin (LJP), 4 place Jussieu, F-75005 Paris, France.}


\begin{abstract}
To mimic the mechanical response of passive biological cilia in complex fluids, we study the bending dynamics of an anchored elastic fiber submitted to a dilute granular suspension under shear.  We show that the bending fluctuations of the fiber accurately encode minute variations of the granular suspension concentration. Indeed, besides the stationary bending induced by the continuous phase flow, the passage of each single particle induces an additional deflection. We demonstrate that  the dominant particle/fiber interaction arises from contacts of the particles with the fiber and we propose a simple elastohydrodynamics model to predict their amplitude. Our results provide a mechanistic and statistical framework to describe particle detection by biological ciliated systems.
\end{abstract}
\maketitle

\maketitle

Living organisms can probe the mechanical features of their immediate environment with exquisite precision and rapidity. Humans for instance, are able to discriminate micrometric variations of the roughness of a solid surface in less than a second, by rubbing their fingertips across it~\cite{FlanaganJohansson2009}. Rodents use their long facial whiskers as tactile organs and have a comparable tactile accuracy for object detection and texture discrimination \cite{arabzadeh2016vibrissal,morita2011psychometric}. Regardless of the specific features of the tactile organ, the early steps of texture detection processes can be sketched into two phases. The first one is mechanical: upon contact and sliding of the tactile organ, the solid texture elicits \emph{mechanical stress fluctuations} which are propagated through the tactile organ. The second phase is biological: mechanosensitive cells embedded in the tactile organ encode the mechanical signal in series of action potentials~\cite{jenkins2017developing,vallbo1984properties,leiser2007responses}. During the first phase, the geometrical and mechanical characteristics of the tactile organ, such as its microstructure and its resonance properties, participate in the texture encoding, by filtering and amplifying rapid fluctuations of the texture induced stress signals~\cite{scheibert2009role,wandersman2011texture,weber2013spatial,boubenec2014amplitude,claverie2017}.\\
\indent However, many living organisms live in aquatic environments. Their tactile detection is thus not mediated by solid friction but rather by hydrodynamic interactions. In many cases, aquatic animals and micro-organisms use myriads of high aspect ratio hair-like or ciliated structures as tactile organs. Their bending in the liquid can trigger the neural response of mechanosensitive cells embedded at their proximity. Fish for instance use their lateral line, a sensory organ made of neuromasts~\cite{montgomery1997lateral,chagnaud2008measuring}, consisting of receptive hair cells that bend under flow and allow fish to detect predators, flow magnitude and direction ~\cite{montgomery1997lateral, oteiza2017novel}. In vertebrates and mammals, the upper surface of the tongue is almost entirely covered with soft slender structures called filiform papillae that can bend when submitted to hydrodynamic flows and trigger the response of mechanosensitive channels at their base~\cite{moayedi}. Filiform papillae are thus likely involved in the in-mouth tactile perception~\cite{Lauga2016,thomazo2019probing}. Because of their ubiquity in Nature, numerous works have studied the deformation of such cilia under given mechanical stresses, both in \textit{in-vivo} \cite{venier1994analysis} and artificial systems \cite{wexler2013bending,chagnaud2008measuring,thomazo2019probing}. For instance, the bending equilibrium of an elastic fiber in a simple steady viscous flow  has been well characterized and accurately modeled at zero Reynolds number using elastohydrodynamics theories~\cite{du2019dynamics,Lauga2016}.\\
\indent Natural biological environments are, however, more complex, often consisting of unsteady flows of heterogeneous media (composed of numerous macromolecules, colloids, granular particles and even other cells). For instance, liquid food products are usually made of oil in water emulsions, and can also contain protein aggregates and solid particles. Yet, Humans can discriminate minute variations of such liquid food texture in the oral cavity. They are able, for instance, to detect the presence of micrometric particles in fluids, even at very low concentrations \cite{engelen,imai}. The implication of filiform papillae in food texture perception is highly suspected but unraveling the precise encoding of food texture by such ciliated structures requires further investigations.\\ 
As for solid texture perception, one can wonder how ciliated structures do encode the texture of a liquid, that one can define \textit{a minima} as the viscosity of the carrier fluid and the typical size and concentration of the surrounding particles. This question remains largely unexplored. In this work, we address it with the use of a minimal biomimetic approach that consists in submitting an anchored elastic cylindrical fiber to a granular suspension flow. We establish both experimentally and theoretically how the fiber bending fluctuations encode the presence of particles in the carrier fluid, as well as their number and their size. We also estimate the typical stresses elicited by particle/fiber interactions in living systems for which the neural responses to mechanical stimuli have been well characterized, such as the fish neuromasts \cite{van1987laser,mchenry2007flexural} and the sense of hearing hair cells~\cite{howard1988compliance}.

\begin{figure*}[t]
\centering
\includegraphics[width=\linewidth]{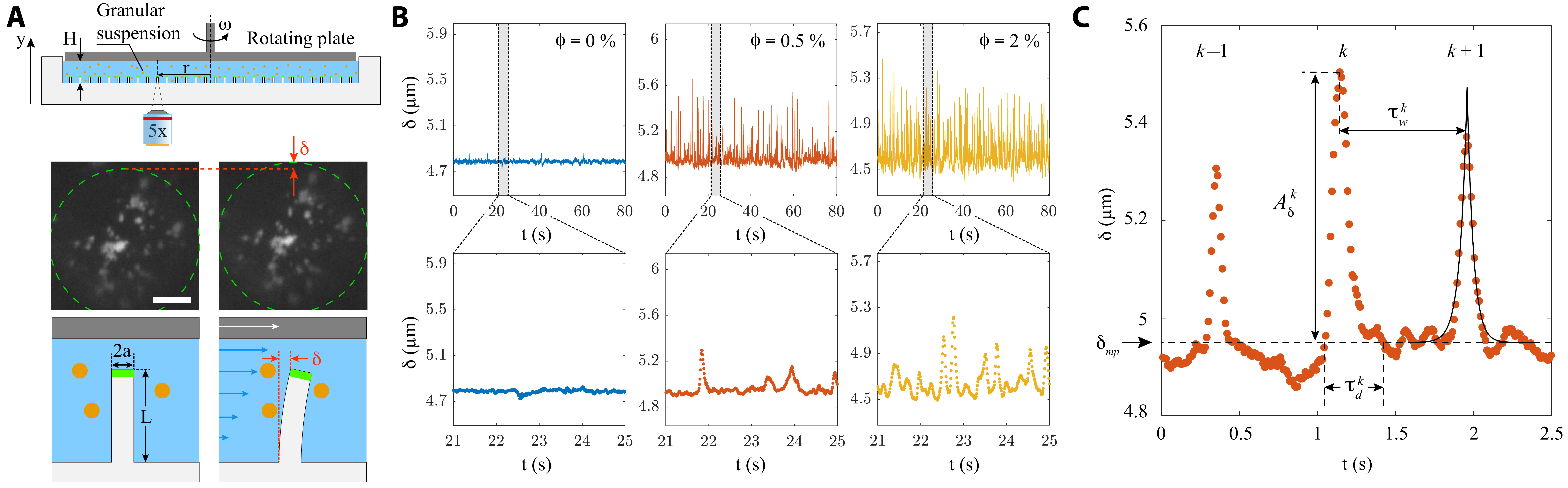}
\caption{(\textit{A}) \textit{Upper panel --} Sketch of the experimental setup. A circular pool made of elastomer, the bottom of which is decorated with cylindrical fibers, is filled with a granular suspension and placed at the static bottom of a rheometer. The rotating tool of the rheometer is rotated to shear the granular suspension. An  epifluorescence microscope, whose objective is placed underneath, images the displacements of fluorescent microparticles embedded at the tip of the fiber. \textit{Middle panel --} Typical fluorescence images of the tip of the fiber, in the absence of flow (\textit{left}) and in steady flow (\textit{right}). \textit{Bottom panel --} Sketch of the geometrical properties of the fiber at rest (\textit{left}) and deformed in a steady flow (\textit{right}). (\textit{B}) Typical time traces of the fiber tip displacements for a particle size $R_0=40~\mu$m with $\phi=0$ (\textit{left}, blue curve), $\phi=0.5 \%$ (\textit{middle}, red curve) and $\phi=2\%$ (\textit{right}, yellow curve). For each trace, a zoom in time (shaded region) is shown just below to highlight a few spike events induced by a particle/fiber interaction. (\textit{C}) Three different spikes labeled $k-1$, $k$ and $k+1$. The spike $k$ has an amplitude $A_\delta^k$ and a duration $\tau_d^k$. The waiting time between spike $k$ and $k+1$ is $\tau_w^k$. The most probable value of the deflection, $\delta_{mp}$, is shown with the dashed line. The solid line is a fit of the $k+1^\mathrm{th}$ event with a symmetric double exponential pulse whose full expression is given in the SI Appendix.}
\label{fig:fig1}
\end{figure*}

\section*{Results}
\subsection*{Fiber displacement}
We monitor optically the deflection of the tip of an elastic cylindrical fiber (radius $a$, height $L$) made of elastomer (Young's modulus $E$, \textit{see} \textit{Materials and Methods} and \textit{SI Appendix}, Fig.~S1). The base of the elastic fiber is anchored to the bottom of a circular pool made of the same elastomer. The pool is filled with a density matched granular suspension consisting of a dispersion of polystyrene spherical particles (radius $R_0=20, 40$ and $70~\mu$m) with a  particle volume fraction $\phi$ (from 0 to 2 \%) in a Polyethylene Glycol (PEG) aqueous solution of viscosity $\eta_0$ (\textit{Materials and Methods} and \textit{SI Appendix}, Fig.~S2B). The suspension is sheared in a rheometer, using a planar circular tool as an upper plate, that rotates at a constant rotation rate $\omega$. Epifluorescence microscopy is used to measure the modulus $\delta$ of the fiber tip displacement over time at 100~frames per second (fps). Our results are all presented in the steady flow regime. The local shear rate of the flow is $\dot{\gamma}=r\omega/H$, where $r$ is the radial position of the fiber with respect to the rotation axis of the rheometer cell and $H$ is the gap distance from the base of the pool to the lower surface of the rotating plate (Fig.~\ref{fig:fig1}\textit{A}, \textit{top panel} and \textit{Materials and Methods} for the flow parameters values). The Reynolds numbers associated to the fiber and to the particles are both small, on the order of $10^{-3}$. The typical relaxation time of the elastic fiber in the viscous flow is $t_{eh}= 4 \pi\eta_0 L^4 / EI (\ln{(2L/a)}+1/2 )$ with $I=\pi a^4/4$ the area moment of inertia of the cylindrical fiber. Its value (here about 1 ms) is small with respect to the flow timescale ($t_{eh}\ll \dot{\gamma}^{-1}$) so that the fiber can be safely modeled in the quasi-static limit~\cite{Lauga2016,young2007stretch}. Biological cilia in natural environments are usually evolving in similar flow regimes~\cite{Lauga2016,mchenry2007flexural,suli2012rheotaxis}.\\  
\par An experiment consists in a 1 minute long sequence of shear. Each experiment is repeated many times ($n=5$ to 20) to have a significant estimate of the variance of $\delta$. Without particles ($\phi$=0), the tip deflection is induced by viscous stresses of the continuous phase flow, which has a Newtonian rheology (\textit{see} \textit{SI Appendix}, Fig.~S2\textit{B}). As we have shown previously in~\cite{thomazo2019probing}, for an anchored fiber sheared in a Newtonian fluid in the steady flow regime, its tip deflection $\delta_0$ is proportional to the local shear stress $\delta_0 = K_0 \frac{L^5}{a^4}\frac{\eta_0 \dot{\gamma}}{E}$. The prefactor $K_0$ has to be determined experimentally as performed in~\cite{thomazo2019probing}.  Due to the variations of the fiber length $L$ and the Young's modulus $E$ from fiber to fiber, its value has to be systematically measured and lies typically between 1 and 5. When a granular suspension ($\phi \ne 0$) is sheared in the system, $\delta(t)$ displays discrete spikes, whose number in a given time interval increases with the particle volume fraction $\phi$ (Fig.~\ref{fig:fig1}\textit{B}).  The following of the paper focuses on the statistical properties of the noise of the displacement measurements and its physical origin.

\subsection*{Bending statistics in granular suspensions}
When a granular suspension is sheared, the signal $\delta(t)$ resembles that of a train of spikes. In the low particle volume fraction limit explored here, each particle passage induces a spike due to the tip displacement of the fiber. Typically, $\delta(t)$ has a baseline signal of amplitude $\delta_{mp}$  on the top of which discrete spikes are overlapped, due to these particle/fiber interactions. Experimentally, we extract $\delta_{mp}$ as the displacement corresponding to the maximum of the probability density function $p(\delta)$ (\textit{see} \textit{SI Appendix}, Fig.~S3). We find that the value of $\delta_{mp}$ is, within experimental error bars, identical to the deflection due to the continuous phase $\delta_0$ (Fig.~S3, insets). Spikes have an amplitude $A_\delta$, whose physical origin and value will be discussed in depth further down, and have a typical time duration $\tau_d$ that we resolve experimentally (Fig.~\ref{fig:fig1}\textit{C}). Since the dispersion of particles is spatially homogeneous, the signal is not time correlated and the autocorrelation of $\delta(t)$ typically decays to zero on a timescale of the order of $\tau_d$ (\textit{see} \textit{SI Appendix}, Fig.~S4\textit{A}). The power spectrum of $\delta(t)$ is flat up to a cutoff frequency of order $1/\tau_d$  (\textit{see} \textit{SI Appendix}, Fig.~S4\textit{B} and C) beyond which it abruptly decays. We have checked both theoretically and experimentally that shear induced migration of the particles did not occur on the timescale of the experiments (\textit{see SI Appendix, Material and Methods}). 

\par We have first measured the variation of the normalized averaged deflection $\left<\delta\right>/\left<\delta_0\right>$  as a function of the particle volume fraction (Fig.~\ref{fig:fig2}, inset), where brackets denote an average over time and over the $n$ repetitions of the experiments, including different fibers and different suspensions. In this range of volume fractions, spikes are well separated and their amplitudes are small with $A_\delta \ll \delta_{mp}$. They thus weakly impact the average, yielding $\left<\delta\right>/\left<\delta_0\right> \approx 1$ at all $\phi$. The large fluctuations of $\left<\delta\right>/\left<\delta_0\right>$ result from sample to sample variations that are likely due to slight variations of the PEG concentration which modifies the viscosity of the continuous phase. To assess this statement, we have used the rheometer (\textit{see} \textit{SI Appendix, Methods}) to measure the macroscopic viscosity of the granular suspension $\eta(\phi)$ normalized by $\eta_0$, its value without particles (open circles in the inset of Fig.~\ref{fig:fig2}). Within sample to sample experimental errors, $\eta/\eta_0$ does not vary significantly with $\phi$ for $\phi<2\%$. Neither the averaged deflection nor the macroscopic viscosity are thus good measurements of the particle presence, concentration or size. In a second step, we have therefore quantified the fiber bending fluctuations as a function of the particle volume fraction. {We plot on Fig.~\ref{fig:fig2} the variance of the deflection, noted $\sigma^2_\delta$, to which we have subtracted the variance in the absence of particles $\sigma^2_0\approx$ 400 nm$^2$, as a function of $\phi$ for all investigated sizes of particles. For a given particle size $R_0$, measuring the bending fluctuations allows to discriminate differences in particle volume fraction of about 0.5\%. Conversely, at a given volume fraction, the bending fluctuations are significantly different for different particle sizes. Note that different pairs of $(R_0,\phi)$ values can yield  the same value of $\sigma^2_\delta-\sigma^2_0$, so that sizes \emph{and} concentrations cannot be discriminated simultaneously. However, since the bending statistics is not gaussian, the associated bending signals are qualitatively and quantitatively different (\textit{see} Fig.~S6 and SI Appendix).

\begin{figure}[!t]
\centering
\includegraphics[width=0.4\textwidth]{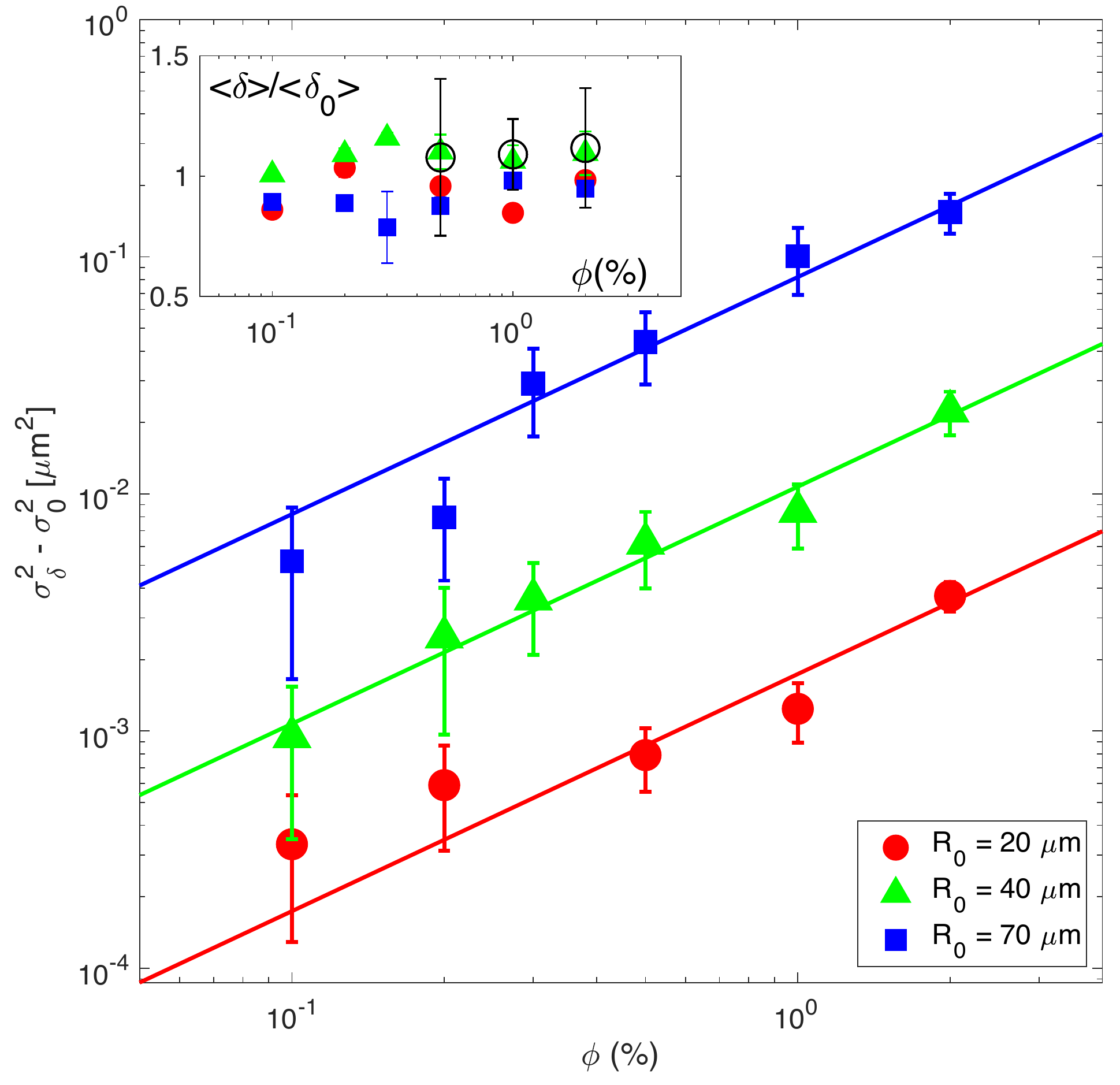}
\caption{Reduced variance of the noise $\sigma^2_\delta-\sigma^2_0$ as a function of $\phi$, with $\sigma_0\approx 20$ nm. Different symbols denote different particle size $R_0$. Solid lines are linear fits of the data $\sigma^2_\delta-\sigma^2_0=\beta(R_0)\phi$.  Inset: Normalized averaged displacement $\left<\delta\right>/\left<\delta_{0}\right>$ as a function of $\phi$, with $\left<\delta_{0}\right>$ the displacement at $\phi=0$. On the same graph we have overlapped the relative increase of the macroscopic viscosity $\eta / \eta_0$ (open circles). Error bars are sample to sample standard deviations of the data.}
\label{fig:fig2}
\end{figure}

\par The individual particle/fiber interactions are sparse and not correlated, therefore one expects the statistics of $\delta(t)$ to be a Poisson process. The anchored fiber may be seen as a particle counter, in a local interaction volume whose cross section will be discussed further down.  One simply expects that the variance of $\delta$ scales linearly with the number of spike events $N(T)$ in a time period $T$, $\sigma^2_\delta\sim N$. Since the number of particles in the interaction volume is proportional to the particle concentration, one thus expects $\sigma_\delta^2\sim \phi$. As shown on Fig.~\ref{fig:fig2} with the solid lines, this is clearly the case for all $R_0$. Data points are well fitted with $\sigma^2_\delta -\sigma^2_0 = \beta(R_0)\phi$, yielding for $\beta(R_0)$ the values given in Table 1.

\begin{figure*}[t]
\centering
\includegraphics[width=0.9\textwidth]{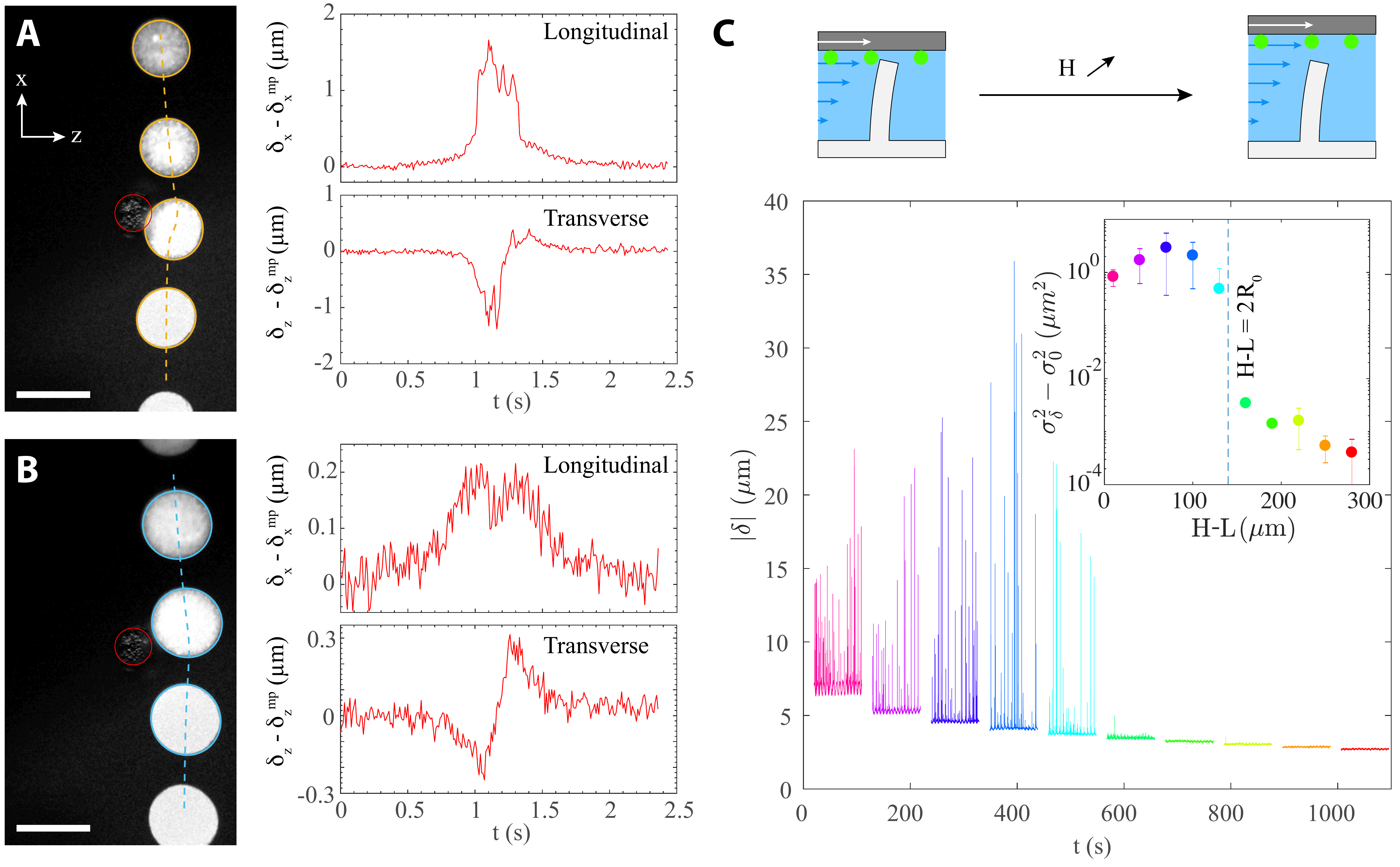}
\caption{A particle interacting with the artificial cilium. (\textit{A}) Snapshots of a typical contact interaction between a fluorescent particle ($R_0=83 \mu$m, circled in orange at different times) and the fiber (circled in red at its steady state position, prior to contact). The trajectory of the particle is shown with the orange dashed line. The white bar is 200~$\mu$m long. On the right of this picture are plotted the corresponding reduced displacements $\delta-\delta_{mp}$ of the tip versus time $t$ in both the direction of the flow $x$ (upper panel) and in the $z$ direction, orthogonal to the flow (lower panel). (\textit{B}) A typical hydrodynamic distant interaction between a fluorescent particle (circled in blue) and the fiber (circled in red). As in (\textit{A}), next to the picture are plotted the corresponding displacements $\delta-\delta_{mp}$ of the tip in both the $x$ (upper panel) and $z$ (lower panel) directions. (\textit{C}) $|\delta|$ versus time $t$ for 10 increasing values of the gap $H-L$, varying from 10~$\mu$m to 280~$\mu$m in increments of 30~$\mu$m. For these experiments, the non-fluorescent particles with the average radius $R_0=70~\mu$m were used. Inset: Reduced  variance $\sigma^2_\delta - \sigma^2_0$ as a function of the gap $H-L$. Error bars correspond to the mean standard deviation of $\sigma^2_\delta - \sigma_0^2$ obtained on 5 successive experiments performed in the same conditions. The vertical dashed line corresponds to $H-L=2R_0$. Colors are the same as in the main panel.} 
\label{fig:fig3}
\end{figure*}

\subsection*{ Individual events: contact \textit{vs.} distant interactions}
To understand the physical nature of the particle/fiber interactions that produce spikes in the tip displacement $\delta$, we imaged the trajectories of the particles. For this, we used fluorescent granular particles ($R_0= 83~\mu$m) to prepare the sheared suspension (\textit{see} \textit{Material and Methods}). Since our optical imaging is restricted to the 2D focal plane, we also unmatched the mass densities of the particles and the liquid, so that particles cream at the top of the suspension, using glycerol as the continuous phase. The rigid plate of the rheometer is lowered so that $H \gtrsim L$. Fluorescent particles therefore approach the fiber at the level of its tip and we extract both the tip deflection (as previously done) and  the particle positions using a tracking routine (\textit{see} \textit{SI Appendix, Material and Methods}). Shown on Figs.~\ref{fig:fig3}\textit{A} and 3\textit{B} are typical snapshots of two interaction events between a particle and the fiber.  On Fig.~\ref{fig:fig3}\textit{A}, we show the case of a contact between the particle and the fiber, for which the particle/fiber distance remains constant during $\tau_d$ (\textit{see} Fig.~S8B and SI Movie S1). Note that at such low Reynolds numbers, this contact remains lubricated. We show on Fig.~\ref{fig:fig3}\textit{B}, a case in which the particle/fiber distance is slightly larger (by about 30 $\mu$m). Movies of both situations can be found in the \textit{SI Appendix} (\textit{SI Movie S1} for a contact and \textit{SI Movie S2} for a distant interaction). Note that since the tips of the fibers are a bit rounded (\textit{see} Fig.~2a in~\cite{thomazo2019probing}) the particle and the fiber can appear to slightly overlap, as in Fig.~\ref{fig:fig3}A.

Next to the snapshots of Figs.~\ref{fig:fig3}\textit{A} and \textit{B} are plotted the corresponding displacements in the flow direction $\delta_x$ and orthogonal to it $\delta_z$. When comparing both figures, one can see that the contact perturbation is about 5 times larger in amplitude than the  distant one. Actually,  distant interactions induce an additional deflection that is hardly detected with our imaging technique, with a typical amplitude close to the noise level in the absence of particles. This is a first indication that the range of particle/fiber interaction is $R_0$, measured from the surface of the fiber.

\begin{table}
\centering

\caption{Comparison between the experimental and predicted values of the coefficient $\beta$ for different particle sizes $R_0$. For the model, numbers given in the brackets correspond to the boundary values of $\beta$ when the length of the fiber $L$ is decreased or increased by  5\% and when $E$ and $\eta$ are varied within their experimental errors.}
		 
		\begin{tabular}{|c|cc|}
		\hline 
		$R_0$~($\mu$m) &  $\beta^{exp}~(\mu$m$^2)$ [Exp.] & $\beta^{th}~(\mu$m$^2)$ [Model] \\ 
		\hline 
		20 &    0.17~$\pm$~0.04 & 0.22 [0.06 ; 0.88] \\ 
		\hline 
		40 &    1.07~$\pm$~0.13 & 0.8 [0.19 ; 2.74]  \\ 
		\hline 
		70 &     8.2~$\pm$~1.2 & 1.96 [0.54 ; 7.74]  \\ 
		\hline 	
		\end{tabular}
\end{table}

\par To further probe the range of these interactions, we performed series of experiments in which the vertical position of the rigid upper plate of the rheometer was changed from $H\sim L$  (\textit{see} Fig.~\ref{fig:fig3}\textit{C} with $t<140$~s) to larger values. Since particles cream in the fluid, as soon as $H>L+2R_0$,  contacts are suppressed and the sole response of distant hydrodynamic interactions is probed (\textit{see} Fig.~\ref{fig:fig3}\textit{C} with $t>560$~s). In this case, $\delta(t)$ is, indeed, much smoother, with a reduced variance $\sigma^2_\delta - \sigma^2_0 \sim 0$ (Fig.~\ref{fig:fig3}\textit{C}, inset). The decay of $\sigma^2_\delta - \sigma^2_0$ typically occurs on a length scale equal to $R_0$ that indeed sets the range of the interaction. Since the amplitudes of the tip deflection induced by contacts are much larger than those due to distant interactions, we focus in the following in developing a model for contacts. A model for distant interactions is also provided and discussed in the \textit{SI Appendix}.

\section*{Theoretical Modeling}
\subsection*{Statistical modeling of the spike train}
The statistical properties of a train of random spikes have been described by Garcia and coworkers \cite{garcia2017auto,garcia2012stochastic}. Following their work, for a duration $T$, $\delta(t)$ can be written as the discrete sum of $N(T)$ random successive spikes as
\begin{equation}
\delta(t) = \delta_{mp} + \sum_{k=1}^{N(T)}A_\delta^k(y,z)\Pi\left(\frac{t-t^k}{\tau_d}\right)
\label{trainspike}
\end{equation}
\noindent where $A_\delta^k(y,z)$ is the random amplitude of the $k^{th}$ spike, $\Pi(x)$ a function describing the shape of the spike (taken here as a symmetric double exponential pulse, \textit{see} Fig.~\ref{fig:fig1}\textit{C} and its full expression in the \textit{SI Appendix}), centered on the time $t_k$ of a given spike, and denoting $\tau_d$ its duration (\textit{see} Fig.~\ref{fig:fig1}\textit{C}). Note that the use of a symmetric double exponential pulse was simply chosen because it fits the experimental spike shape and allows for a simple computation of the normalization integral $I_2$ of Eq.~\ref{Eqdmoy} defined further down. This choice is not supported by any physical model. Both the spike duration $\tau_d$ and its amplitude $A_\delta^k$ are random variables that depend on the random $(y_0,z_0)$ coordinates of the granular particles that interact with the fiber (\textit{see} the sketch of Fig.~\ref{fig:fig4}\textit{A}). Denoting $\left<\tau_w\right>$ the average waiting time between spikes, the variance of $\delta$ can be derived, by computing the long time averaging of the square of Eq.~\ref{trainspike}, under both assumptions that the spikes are independent and that they are well separated $\tau_d\ll \tau_w$. One obtains   
\begin{equation}
\sigma_\delta^2 =\frac{\left<A^2_\delta\tau_d\right>}{\left<\tau_w\right>} I_2
\label{Eqdmoy}
\end{equation}
\noindent where the average is taken on the particle coordinates disorder and $I_2 = \int_{-\infty}^{\infty}\Pi^2(x)dx=1/2$ is a shape normalization integral. The average waiting time between spikes $\left<\tau_w\right>$ can be estimated from the flux of particles that effectively interact with the fiber, $\tau_w=V_p/(\phi L^2\dot{\gamma}b)$ with $V_p=4/3 \pi R_0^3$ the particle volume and $b=a+R_0$ the characteristic impact parameter length of the particle/fiber interaction. We now provide a physical estimate of $\left<A_\delta^2 \tau_d\right>$ for contact interactions. 

\subsection*{Model for the particle/fiber contact}
In the case of a contact, we assume that the fiber experiences an enhanced drag force due to the presence of the particle in its immediate vicinity, responsible for the additional bending of the fiber. We model this force as a simple Stokes drag \textbf{F}$(y_0,z_0)=-6\pi \eta_0 R_0 \textbf{u}$, where $\textbf{u}$ is taken as the local fluid velocity around a cylindrical obstacle (\textit{see SI Appendix}) and where $y_0$  (\textit{resp.} $z_0$) is the vertical (\textit{resp.} lateral) coordinate of the particle (\textit{see} the sketch of Fig.~\ref{fig:fig4}\textit{A}). Since the elastohydrodynamic relaxation time of the fiber ($t_{eh}\sim$1~ms) is much smaller than the interaction duration ($\tau_d\sim$0.1~s), the bending dynamics is not governed by the relaxation timescale of the fiber and the additional deflection can be modeled in the quasi-static limit. For the sake of simplicity, we only keep the radial component of the force, along the particle/fiber center-to-center direction. Using linear elasticity theory, the fiber profile $\delta(y)$ can be solved, writing that $\delta''(y) = \frac{F(y_0,z_0)}{E I}(y_0-y)$, where the prime symbol stands for a spatial $y$ derivative and $I=\pi a^4/4$ is the area moment of inertia of the fiber. Considering the boundary conditions (fiber clamped at its base so that $\delta(0)=0$ and $\delta'(0)=0$, pinned at the contact point so that $\delta'(y_0^-)=\delta'(y_0^+)$ and free at its tip so that $\delta''(y>y_0)=0$), we obtain the following expression for the amplitude $A_\delta$

\begin{equation}
\footnotesize
A_\delta=\frac{4\eta \dot{\gamma}}{E} \left(1-\frac{a^2}{(a+R_0)^2}\right) \frac{R_0 y_0^3(3L-y_0)}{a^4} \sqrt{1-\left(\frac{z_0}{R_0+a}\right)^2}
\label{Eq:Ad}
\normalsize
\end{equation}

\noindent Looking at the experiments with fluorescent granular particles, we observe that during a contact event, the particle encounters the fiber at a lateral position $z_0$, which is a random variable. In the fiber frame, it corresponds to an angular coordinate $\theta_i$ (\textit{see SI Appendix} Fig.~S8A). The particle detaches from the fiber at an angle $\theta_f$. Experimentally, we found that $\theta_f$ weakly depends on $\theta_i$, with $<\theta_f> = 108\pm 4^\circ$  (\textit{see SI Appendix} Fig.~S8C). We make the crude assumption that the particle travels a distance $(a+R_0)(\theta_f-\theta_i)$, at the unperturbed flow velocity $\dot{\gamma}y_0$, yielding $\tau_d  =  \frac{\theta_f(a+R_0)-z_0}{\dot{\gamma}y_0}$. We can thus average out $A_\delta^2 \tau_d$ over space, taking into account a uniform distribution for the $(y_0,z_0)$ particle coordinates. After all computations have been made, one finally gets (\textit{see} the full calculation in the \textit{SI Appendix, Theoretical Modeling})
\begin{equation}
\left(\frac{\sigma_{\delta}}{L}\right)^2  = \frac{4\theta_f}{\pi}\left(\frac{\eta\dot{\gamma}}{E}\right)^2\left( \frac{L}{a}\right)^8\left(\frac{R_0+2a}{R_0+a}\right)^2 \times \mathcal{P}_2(R_0/L)  \times \phi 
\label{Eqmodel}
\end{equation}
\noindent with~$\mathcal{P}_2(x) = 43/56x+4x^2+8x^3+7x^4+5/4 x^5-2x^6-5/2x^7+x^8$.\\

\par Comparison of this model's predictions with the experimental data was done in the following way. First, we looked at an individual contact event with the fluorescent particles data and compared the model prediction for $A_\delta(y_0\approx L,z_0)$, as provided by Eq.~\ref{Eq:Ad}, to the experiments. For moderate deflections ($A_\delta < \delta_{mp}$), we find that our model is in reasonable agreement with the data both for $A_\delta$ and $\tau_d$ (\textit{see} \textit{SI Appendix}, Fig.~S8D).  We then confronted our model to the dataset of different particle concentrations and sizes (density matched suspensions). As we mentioned above, the signal $\delta(t)$ is that of a particle counter, which explains the scaling $\sigma^2_\delta \sim \phi$ with the particle concentration $\phi$. The standard deviation is also expected to scale linearly with the strain rate, in agreement with the experimental data (\textit{see} \textit{SI Appendix}, Fig.~S7B). We have also compared the experimental values of $\beta/L^2$ (\textit{see} Fig.~2) to its predicted ones as obtained from Eq.~\ref{Eqmodel} and given in Table 1.  As shown on Fig.~\ref{fig:fig4}\textit{B}, within experimental error bars, and without any adjustable parameters, our model describes well the variation of $\beta$ with the particle size $R_0$.  Averaging over the three particle sizes, we find $\left<\beta^{exp}/\beta^{th}\right>\approx 2.1$, which is in the range of the measured values of the calibration factor $K_0$ in a shear flow without particles. 

\begin{figure}[h]
\centering
\includegraphics[width=0.4\textwidth]{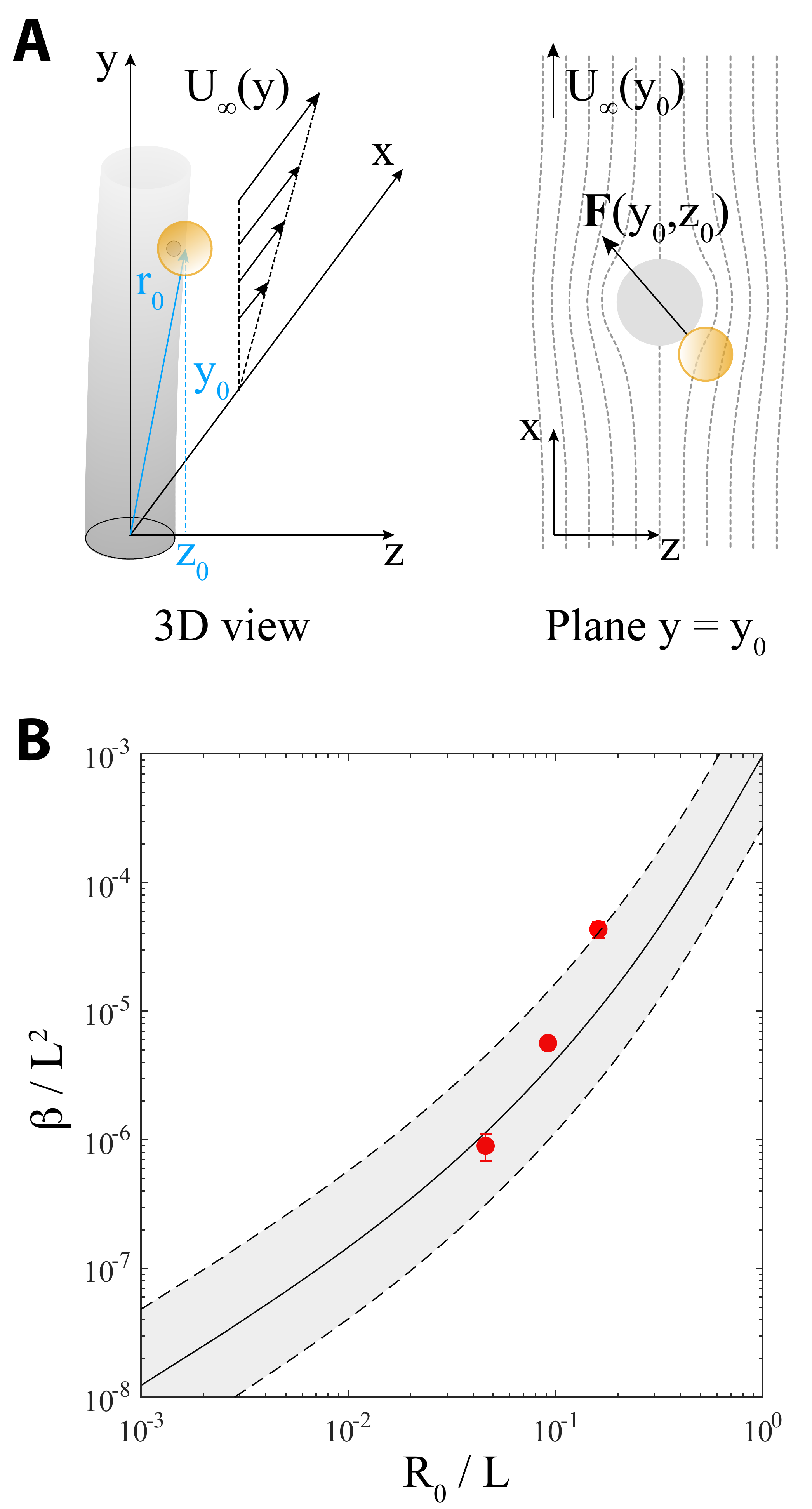}
\caption{(\textit{A}) Sketch of the physical situation for a particle/fiber contact interaction. \textit{Left panel} -- 3D view. \textit{Right panel} -- Top view in the plane $y=y_0$. $U_\infty$ denotes the flow magnitude far away from the fiber. (\textit{B}) Red disks correspond to the experimental values of $\beta$ (\textit{see} Table 1) divided by $L^2$ as a function of the normalized particle size $R_0/L$ .The solid line corresponds to the theoretical predictions of the contact model, with $a=50~\mu$m, $L=435~\mu$m, $E=2.7$~MPa and $\eta=119.7$~mPa.s. The dashed lines estimate the model prediction limits taking $\pm5\%$ variations in $L$ and variations of $E$ and $\eta$ within their experimental errors. }
\label{fig:fig4}
\end{figure}

\section*{Discussion}
Several conclusions can be drawn from our results. On a physical side, the increase of particle volume fraction leads, at the macroscopic scale, to an increase of the shear viscosity, according to Einstein's law for suspensions: $\eta / \eta_0 = 1+5/2\phi+o(\phi^2)$. This relationship has been verified experimentally using standard rheological methods \cite{brodnyan1968dependence,chong1971rheology}, but mostly at particle volume fractions larger than a few \%. Detecting a particle volume fraction of particles lower than 1\% using a viscosity measurement would indeed require a relative error on $\eta$ better than $\sim 2\%$. Such a small error is beyond current rheometry experimental errors, due to rheological artefacts such as gap detection \cite{davies2005gap} or the presence of a rim at the edge of the sample \cite{cardinaels2019quantifying}, which both lead typically to relative errors on the order of 10\%. \\
Actually, to discriminate the macroscopic viscosities of the suspensions, the biomimetic cilium is not more efficient than a rheometer, since it does not capture the `mean field' increase of the macroscopic viscosity. Within our experimental error bars, the average deflection does not vary significantly with the particle volume fraction $\phi$. However, the biomimetic cilium is very sensitive to the local environment heterogeneity through repeated particle/fiber interactions. The interaction between an anchored elastic fiber and granular particles has already been explored, but mostly with \emph{active} ciliated assemblies adhering to particles in suspensions~\cite{shields2010biomimetic,zhang2019removal}, to mimic the beating of some biological cilia, for instance in the context of mucus cleaning in the lungs. In~\cite{bhattacharya2012propulsion}, the authors show that by varying the  adhesion strength, particles can either be released, propelled, or trapped by cilia. Surprisingly, the apparently simpler case of a non-adhesive particle interacting with a passive and isolated cilium has been less investigated. Non-adhesive particle/cilia interactions were probed in \cite{mikulich2015experimental} but in the limit $R_0 \gg a$. Severe deformations of the pillar assemblies are observed, that could lead to the rupture of the cilia.  Our work contrasts with these results by focusing on the case of non-adhesive particles with $R_0 \sim a$, interacting with an isolated cilium.\\

\par In living systems, mechanoreceptors are usually located at the base of the cilia and one can question whether these individual particle/cilium interactions induce sufficient mechanical signals to trigger a neural and behavioral response. Humans are capable of detecting with their tongues the presence of rigid particles in fluids with sizes as low as 2 $\mu$m and concentrations as low as 5\% \cite{engelen}. This detection was experimentally found to depend on the size of the particles, their concentration and the carrier liquid viscosity~\cite{imai}. These behavioral experiments tend to show that particles do induce a neural response, even though it has not been explicitly measured.
Usually, the neural response threshold of ciliated structures is obtained by indenting the tip of the cilia while recording the activity of basal mechanoreceptors. This response threshold corresponds to a torque at the base of the cilium, that has to be compared here with the typical torque induced by a cilium/particle contact.\\
\indent In the case of a contact between a single particle and a cilium, the base torque $M_b$ can be deduced from the cilium base curvature and is given by $M_b=EI \delta''(y=0)=F y_0$, where $F$ is the Stokes viscous force mentioned above. For multiple interactions, an average torque $\left<M_b\right>$ can be computed by averaging over all possible particle positions (\textit{see} the full derivation in the \textit{SI Appendix}). Assuming a uniform flow of velocity $U_\infty$ and for particles of size $R_0\approx a$, one finds that 
    \begin{equation}
    \left<M_b\right>\approx \frac{9}{16} \pi^2 \eta U_\infty a (L+2a)
    \label{BaseTork}
    \end{equation}
\indent Measurements of base torques in ciliated living systems are rather scarce. One of the most thoroughly studied ciliated system is the lateral line of the Zebrafish larvae, which is composed of superficial neuromasts. These can be assimilated to slender cylindrical structures of typical length $L\approx40~\mu$m and base diameter $2a\approx8~\mu$m. Using either laser interferometry or direct microfiber indentation, a typical bending base torque of about $3.10^{-15}$~N.m was estimated~\cite{van1987laser,mchenry2007flexural}. How does this value compare to the mean base torque induced by particle/neuromasts interactions~? For a larvae swimming in water (viscosity $\eta\approx 1$~mPa.s) with a typical velocity $U_\infty\sim10^{-2}$~m/s~\cite{GokulThesis}, Eq.~\ref{BaseTork} yields $\left<M_b\right>\approx 10^{-14}$~N.m, a value that compares well with the above estimation. Could such a base torque value be sufficient to induce a neural response from the neuromast afferent fibers~? To the best of our knowledge, on Zebrafish larvae, no measurements that combine torque estimates together with electrophysiology measurements were actually provided. Such combined measurements have in fact been performed by Hudspeth and coworkers in another, yet closely related system, the hair bundle of the Bullfrog saccular hair cell \cite{howard1988compliance} that has similar sizes and compliance as the neuromast. In their work, Hudspeth and coworkers show that the onset of the neural response is triggered for forces acting on the bundle $F \approx 10^{-11}$~N. Since the bundle has a typical height of 10~$\mu$m, this force yields a threshold base torque $M_b\approx 10^{-16}$~N.m that is smaller than our predicted value.\\ 
\indent This strongly suggests that a single contact interaction between the cilium and micrometer sized particles could therefore be detectable by the lateral line of fish or any comparable hair cell structures. It is known for instance that cilia are involved in the left/right symmetry breaking during embryogenesis, possibly \emph{via} the detection of morphogens filled vesicles~\cite{daems2020fluid}. For a ciliated biological system to measure particle concentrations, a statistical measurement over time of several of these individual events is required. Our results show that the variance of the bending signal could be used by biological systems to discriminate sizes or concentrations. In biological systems, it is well known that sensory systems also encode stimuli with the precise timing of the events~\cite{FlanaganJohansson2009}. Encoding simultaneously \emph{both} sizes and concentrations therefore calls for a more refined statistical analysis and a direct confrontation with biological systems. For instance, one could image the fluctuating deflections of a cilium under the flow of a colloidal suspension at various concentrations and perform at the same time electrophysiological measurements of its afferent neurons.  
\section*{Materials and Methods}
\subsection*{Fiber fabrication}
Elastic fibers were obtained using micro-milling and molding techniques fully described in~\cite{thomazo2019probing} and recalled in the \textit{SI Appendix, Material and Methods}. They are made of a PolyDiMethylSiloxane elastomer (PDMS, Sylgard 184, Dow Corning, USA, cross-linking ratio 10:1, Young's modulus $E \approx 2.7\pm0.8$~MPa) and consist in cylinders (height $L$=435$\pm20\mu$m, radius $a$=50 $\mu$m) whose base is anchored to a circular pool made of the same elastomer. To image their displacements with fluorescence microscopy, their tips were seeded with fluorescent microspheres (diameter $\sim$ 5 $\mu$m) using the protocol described in~\cite{thomazo2019probing}. 
\subsection*{Granular suspensions}
Two types of granular suspensions were used. The first one consisted in a suspension of Polystyrene spherical particles (TS-40, TS-80 and TS-140 of mean radius $R_0=20, 40$ and 70 $\mu$m respectively (Fig.~S2A), mass density $1.05$~g/cm$^3$, Dynoseeds, Microbeads, Norway) dispersed in a dilute solution of Polyethylene Glycol (PEG, $M_w=8.10^3$~g.mol$^{-1}$, Sigma Aldrich) with a particle volume fraction $\phi$ ranging from 0 to 2 \%. For these suspensions, a PEG mass concentration of 30\% w/w was used. This allowed having a solution with a rheology that is still Newtonian (dynamic viscosity $\eta \approx 130$~mPa.s, \textit{see} \textit{SI Appendix}, Fig.~S2B) and with a mass density that closely matches that of the particles, thus limiting their sedimentation or creaming. The second type of suspension consisted of green fluorescent Polyethylene particles (mean radius $R_0$=83 $\mu$m, UVPMS-BG-1.025, mass density $1.025$~g/cm$^3$, $\lambda_{em} = 414$~nm, Cospheric, USA) dispersed in pure glycerol (mass density $1.26$~g/cm$^3$, Sigma Aldrich) with $\phi\ll1$ \%. In this case, the fluorescent particles did cream in the granular suspension.
\subsection*{Rheological and optical setup}
The PDMS circular pool was placed at the bottom static part of a commercial rheometer (MCR 302, Anton Parr). The planar rotating tool of the rheometer (PP40 Anton Paar, diameter 40 mm) was positioned at a height $H$=1~mm above the base of the fiber and rotated at a constant rotation rate (yielding $\dot{\gamma}$=10 Hz for all experiments with non fluorescent particles) to induce a shear flow  (\textit{Fig.~1A}). For fluorescent particles experiments the rheometer tool was positioned at a height $H$=540~$\mu$m above the base and the constant shear rate was in the range [-2 ; 2]~Hz. A fluorescence microscope equipped with a 5X magnification objective is positioned underneath the pool. Images of the tip position were recorded at 100 fps as the granular suspension is sheared with a CCD sensitive Camera (Blackfly S, FLIR, USA). \\
\subsection*{Fiber deflections measurements}
We used an image correlation routine written in Matlab (Mathworks, USA) to compute the displacement of the fiber tip $\delta_x$ (in the the flow direction) and $\delta_z$ (orthogonal to the flow). Sub-pixel accuracy is obtained by interpolating the correlation function as fully described in \cite{thomazo2019probing}, yielding a typical measurement noise $\sigma_0=20$~nm.
}
\begin{acknowledgments}
The authors wish to acknowledge financial support from Centre de Recherche et D\'eveloppement Nestl\'e S.A.S., Marne la Vall\'ee, France, and Société des Produits Nestlé SA, Vevey, Switzerland. They also wish to thank Isabelle Barbotteau and Giulia Marchesini (Nestl\'e Development Centre Lisieux, France) for their careful reading of the manuscript and support. The authors acknowledge as well the continuous support of Christopher J. Pipe (Nestl\'{e} Research, Switzerland). Finally, the authors wish to thank Eric Lauga (DAMTP, Cambridge Univ., UK), Georges Debr\'egeas (Laboratoire Jean Perrin, Paris, France) and Christian Fr\'etigny (Laboratoire SIMM, Paris, France) for fruitful discussions.
\end{acknowledgments}

%

\end{document}